\pdfoutput=1
\documentclass[10pt,conference]{IEEEtran}

\usepackage{xspace}
\usepackage{booktabs}   
\usepackage{subcaption} 
\usepackage{xcolor}
\usepackage{textcomp}
\usepackage{pgfpages}
\usepackage{pgfplots}
\usepackage{tikz}
\usepackage{enumerate}
\usepackage{algorithmicx}
\usepackage{algorithm}
\usepackage[noend]{algpseudocode}
\usepackage{multirow}
\usepackage{amsmath}
\usepackage{amsthm}
\usepackage{listings}
\usepackage{colortbl}
\usepackage{url}
\usepackage{titlecaps}
\usepackage[font=bf]{caption}
\usepackage{flushend}

\usepackage{hyperref}

\hypersetup{breaklinks,hidelinks,pdftitle={Targeted Greybox Fuzzing with Static Lookahead Analysis},pdfauthor={Valentin Wuestholz,
Maria Christakis},pdfkeywords={greybox fuzzing, dynamic program
analysis, static program analysis}}
\definecolor{gray}{rgb}{0.5, 0.5, 0.5}
\definecolor{light-gray}{gray}{0.77}
\definecolor{BrickRed}{rgb}{0.8, 0.25, 0.33}
\definecolor{Black}{rgb}{0.0, 0.0, 0.0}
\definecolor{DarkBlue}{rgb}{0.0, 0.0, 0.55}
\definecolor{Crimson}{rgb}{0.86, 0.08, 0.24}
\definecolor{SlateGrey}{rgb}{0.44, 0.5, 0.56}
\definecolor{lightorange}{HTML}{FFB74D}
\definecolor{blue}{rgb}{0.0, 0.0, 1.0}
\definecolor{magenta}{rgb}{0.79, 0.08, 0.48}

\makeatletter
\newenvironment{btHighlight}[1][]
{\begingroup\tikzset{bt@Highlight@par/.style={#1}}\begin{lrbox}{\@tempboxa}}
{\end{lrbox}\bt@HL@box[bt@Highlight@par]{\@tempboxa}\endgroup}

\newcommand\btHL[1][]{%
  \begin{btHighlight}[#1]\bgroup\aftergroup\bt@HL@endenv%
}
\def\bt@HL@endenv{%
  \end{btHighlight}%
  \egroup
}
\newcommand{\bt@HL@box}[2][]{%
  \tikz[#1]{%
    \pgfpathrectangle{\pgfpoint{0.3pt}{0pt}}{\pgfpoint{\wd #2}{\ht #2}}%
    \pgfusepath{use as bounding box}%
    \node[anchor=base west,fill=lightorange,outer sep=0pt,inner xsep=0.3pt,inner ysep=0pt,minimum height=\ht\strutbox+0.3pt,#1]{\raisebox{0.3pt}{\strut}\strut\usebox{#2}};
  }%
}
\makeatother

\makeatletter
\newif\if@anonymize

\@anonymizetrue

\if@anonymize
  \newcommand\anonymize[1]{Link removed for double-blind review.}
\else
  \newcommand\anonymize[1]{\tiny#1}
\fi
\makeatother

\usetikzlibrary{calc,trees,positioning,arrows,chains,shapes.geometric,%
  decorations.pathreplacing,decorations.pathmorphing,decorations.text,shapes,%
  matrix,shapes.symbols,patterns,shadows,automata}

\def\addlegendimage{\csname pgfplots@addlegendimage\endcsname}

\pgfplotsset{compat=newest}

\newcommand*\LRet[1]{\State \textbf{return} #1}
\newcommand*\Let[2]{\State #1 $\gets$ #2}
\newcommand*\LetHL[2]{\State {\btHL[fill=light-gray]#1 $\gets$ #2}}
\newcommand*\Fcall[1]{\textsc{#1}}

\algrenewcommand\alglinenumber[1]{\tiny\color{Black!70}{#1}}
\algrenewcommand\algorithmicforall[2]{\textbf{for} $i=$ #1 \textbf{to} #2}
\algrenewtext{ForAll}{\algorithmicforall}
\algnewcommand\algorithmicswitch{\textbf{switch}}
\algnewcommand\algorithmiccase{\textbf{case}}
\algdef{SE}[SWITCH]{Switch}{EndSwitch}[1]{\algorithmicswitch\ #1\ \algorithmicdo}{\algorithmicend\ \algorithmicswitch}%
\algdef{SE}[CASE]{Case}{EndCase}[1]{\algorithmiccase\ #1}{\algorithmicend\ \algorithmiccase}%
\algtext*{EndSwitch}%
\algtext*{EndCase}%
\algdef{SE}[SUBALG]{Indent}{EndIndent}{}{\algorithmicend\ }%
\algtext*{Indent}
\algtext*{EndIndent}

\definecolor{verylightgray}{rgb}{.97,.97,.97}

\lstdefinelanguage{Solidity}{
	keywords=[1]{anonymous, assembly, assert, balance, break, call, callcode, case, catch, class, constant, constructor, continue, contract, debugger, default, delegatecall, delete, do, else, event, export, external, false, finally, for, function, gas, if, implements, import, in, indexed, instanceof, interface, internal, is, length, library, log0, log1, log2, log3, log4, memory, modifier, new, payable, pragma, private, protected, public, pure, push, require, return, returns, revert, selfdestruct, send, storage, struct, suicide, super, switch, then, this, throw, transfer, true, try, typeof, using, value, view, while, with, addmod, ecrecover, keccak256, mulmod, ripemd160, sha256, sha3}, 
	keywordstyle=[1]\color{DarkBlue}\bfseries,
	keywords=[2]{address, bool, byte, bytes, bytes1, bytes2, bytes3, bytes4, bytes5, bytes6, bytes7, bytes8, bytes9, bytes10, bytes11, bytes12, bytes13, bytes14, bytes15, bytes16, bytes17, bytes18, bytes19, bytes20, bytes21, bytes22, bytes23, bytes24, bytes25, bytes26, bytes27, bytes28, bytes29, bytes30, bytes31, bytes32, enum, int, int8, int16, int24, int32, int40, int48, int56, int64, int72, int80, int88, int96, int104, int112, int120, int128, int136, int144, int152, int160, int168, int176, int184, int192, int200, int208, int216, int224, int232, int240, int248, int256, mapping, string, uint, uint8, uint16, uint24, uint32, uint40, uint48, uint56, uint64, uint72, uint80, uint88, uint96, uint104, uint112, uint120, uint128, uint136, uint144, uint152, uint160, uint168, uint176, uint184, uint192, uint200, uint208, uint216, uint224, uint232, uint240, uint248, uint256, var, void, ether, finney, szabo, wei, days, hours, minutes, seconds, weeks, years},	
	keywordstyle=[2]\color{teal}\bfseries,
	keywords=[3]{block, blockhash, coinbase, difficulty, gaslimit, number, timestamp, msg, data, gas, sender, sig, value, now, tx, gasprice, origin},	
	keywordstyle=[3]\color{violet}\bfseries,
        keywords=[4]{minimize},
        keywordstyle=[4]\color{BrickRed}\bfseries,
	identifierstyle=\color{black},
	sensitive=false,
	comment=[l]{//},
	morecomment=[s]{/*}{*/},
	commentstyle=\color{purple}\ttfamily,
	stringstyle=\color{red}\ttfamily,
	morestring=[b]',
	morestring=[b]"
}

\lstdefinestyle{solidity}{
	language=Solidity,
	extendedchars=true,
        upquote          = true,%
	basicstyle=\footnotesize\ttfamily,
        columns          = [c]fixed,%
        aboveskip        = 0mm,%
        belowskip        = 2mm,%
        keepspaces       = true,%
        mathescape       = true,%
	showstringspaces=false,
	showspaces=false,
	numbers=left,
	numberstyle=\tiny\color{Black!70},
	numbersep=4pt,
	tabsize=2,
	breaklines=true,
	showtabs=false,
	captionpos=b,
        escapechar=¤,
        moredelim=**[is][{\btHL[fill=light-gray]}]{°}{°},
        xleftmargin=1.3em,%
}

\lstdefinestyle{basic}{%
  morekeywords     = [1]{var},%
  morekeywords     = [2]{},%
  keywordstyle     = [2]\color{teal}\bfseries,%
  morekeywords     = [3]{minimize},%
  keywordstyle     = [3]\color{BrickRed}\bfseries,%
  keywordstyle     = \bfseries\color{DarkBlue},%
  commentstyle     = \ttfamily\color{Black!70}\lst@ifdisplaystyle\footnotesize\fi,%
  basicstyle       = \ttfamily\lst@ifdisplaystyle\footnotesize\fi,%
  emph             = {int,char,double,float,unsigned,void,bool},%
  emphstyle        = {\color{teal}\bfseries},%
  columns          = [c]fixed,%
  aboveskip        = 0mm,%
  belowskip        = 2mm,%
  keepspaces       = true,%
  mathescape       = true,%
  escapechar       = ¤,%
  tabsize          = 2,%
  numbers          = left,%
  numberstyle      = \tiny\color{Black!70},%
  numbersep        = 4pt,%
  stepnumber       = 1,%
  firstnumber      = 1,%
  showstringspaces = false,%
  captionpos       = b,%
  extendedchars    = true,%
  upquote          = true,%
  abovecaptionskip = 0mm,%
  belowcaptionskip = 0mm,%
  moredelim        = **[is][{\btHL[fill=light-gray]}]{°}{°},%
}

\lstdefinestyle{clang}{%
  language         = C,%
  style            = basic,%
}

\newcommand\code[1]{\lstinline[style=clang]{#1}}

\newcommand\harvey{\textsc{Harvey}\xspace}
\newcommand\bran{\textsc{Bran}\xspace}

\newtheoremstyle{mydefinition}
  {}
  {}
  {}
  {}
  {\itshape}
  {.}
  { }
  {}%

\theoremstyle{mydefinition}
\newtheorem{definition}{Definition}

\begin{document}

\bstctlcite{IEEEexample:BSTcontrol}

\title{Targeted Greybox Fuzzing\\ with Static Lookahead Analysis}

\author{\IEEEauthorblockN{Valentin W{\"{u}}stholz}
\IEEEauthorblockA{\textit{ConsenSys Diligence, Germany} \\
valentin.wustholz@consensys.net}
\and
\IEEEauthorblockN{Maria Christakis}
\IEEEauthorblockA{\textit{MPI-SWS, Germany} \\
maria@mpi-sws.org}
}

\maketitle

\begin{abstract}
  Automatic test generation typically aims to generate inputs that
  explore new paths in the program under test in order to find
  bugs. Existing work has, therefore, focused on guiding the
  exploration toward program parts that are more likely to contain
  bugs by using an offline static analysis.

  In this paper, we introduce a novel technique for targeted greybox
  fuzzing using an \emph{online} static analysis that guides the
  fuzzer toward a set of \emph{target locations}, for instance,
  located in recently modified parts of the program. This is achieved
  by first \emph{semantically} analyzing each program path that is
  explored by an input in the fuzzer's test suite. The results of this
  analysis are then used to control the fuzzer's specialized power
  schedule, which determines how often to fuzz inputs from the test
  suite. We implemented our technique by extending a state-of-the-art,
  industrial fuzzer for Ethereum smart contracts and evaluate its
  effectiveness on 27 real-world benchmarks. Using an online analysis
  is particularly suitable for the domain of smart contracts since it
  does not require any code instrumentation---adding instrumentation
  to contracts changes their semantics. Our experiments show that
  targeted fuzzing significantly outperforms standard greybox fuzzing
  for reaching 83\% of the challenging target locations (up to 14x of
  median speed-up).
\end{abstract}





\section{Introduction}
\label{sect:intro}

Automatic test generation is known to help find bugs and security
vulnerabilities, and therefore, improve software quality. As a result,
there has emerged a wide variety of test-generation tools that
implement techniques such as random
testing~\cite{ClaessenHughes2000,CsallnerSmaragdakis2004,PachecoLahiri2007}
and blackbox fuzzing~\cite{PeachFuzzer,ZzufFuzzer}, greybox
fuzzing~\cite{AFL,LibFuzzer} as well as dynamic symbolic
execution~\cite{GodefroidKlarlund2005,CadarEngler2005} and whitebox
fuzzing~\cite{GodefroidLevin2008,CadarDunbar2008,GaneshLeek2009}.

These techniques differ from each other in how much of the program
structure they take into account. In general, the more structure a
testing tool may leverage, the more effective it becomes in
discovering new paths, but the less efficient it is in generating new
inputs. For example, greybox fuzzing lies in the middle of this
spectrum between performance and effectiveness in increasing coverage.
In particular, it uses lightweight runtime monitoring that
makes it possible to distinguish different paths, but it may not
access any additional information about the program under test.

What these techniques have in common is that, just like any (static or
dynamic) path-based program analysis, they can usually only explore a
subset of all feasible paths in a program under test; for instance, in
the presence of input-dependent loops.
For this reason, path-based program analyses are typically not able to
prove the absence of errors in a program, only their existence.

To make bug detection more effective, existing work has focused on
guiding the exploration toward warnings reported by a static analysis
(e.g.,~\cite{CsallnerSmaragdakis2005,GeTaneja2011,CzechJakobs2015}),
unverified program executions
(e.g.,~\cite{ChristakisMueller2016,FerlesWuestholz2017}), or sets of
dangerous program locations (e.g.,~\cite{BoehmePham2017}).
The motivation behind these approaches is to identify safe program
paths at compile time and avoid them at runtime. This is often
achieved with an offline static analysis whose results are recorded
and used to prune parts of the search space that is then
explored by test generation.

The offline static analysis may be semantic, e.g., based on abstract
interpretation, or not, e.g., based on the program text or its
control-flow graph.
A semantic analysis must consider all possible program inputs and
states in which a piece of code may be executed. As a result, the
analysis can quickly become imprecise, thus impeding its purpose of
pruning as much of the search space as possible. For better results,
one could resort to a more precise analysis, which would be less
efficient, or to a more unsound analysis. The latter would limit the
number of considered execution states in order to increase precision,
but may also prune paths that are unsoundly
verified~\cite{LivshitsSridharan2015}.

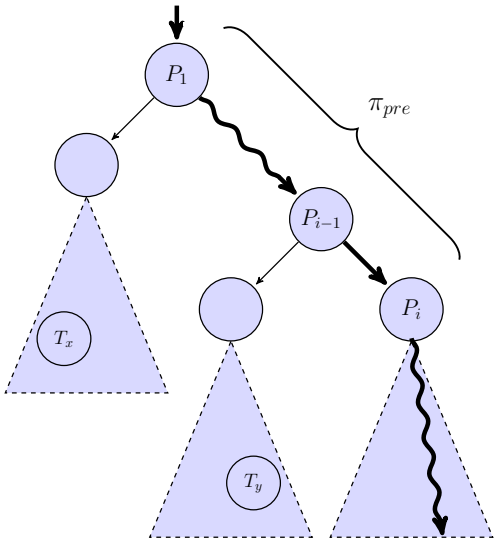
\begin{figure}[t]
\vspace{-2em}
\center
\scalebox{0.6}{
\begin{tikzpicture}[->,>=stealth',shorten >=2pt,auto,node distance=2cm,
  thick,main node/.style={circle,fill=blue!15,draw,
  font=\sffamily\Large\bfseries,minimum width=14mm}]

  \node[main node,draw=white,fill=white] (P0) {~~};
  \node[main node] (P1) [below of=P0,yshift=-0.25cm] {$P_1$};
  \node[main node] (Pbr) [below of=P1, right of=P1,yshift=-1.2cm,xshift=1.2cm]  {$P_{i-1}$};
  \node[main node] (Pi) [below of=Pbr, right of=Pbr] {$P_i$};
  \node[main node] (Pj) [below of=Pbr, left of=Pbr] {~~};
  \node[main node] (Pk) [below of=P1, left of=P1] {~~};
  \node[main node,style={draw,dashed,shape border uses incircle,isosceles triangle,shape border rotate=90,yshift=-0.25cm,scale=1.8,minimum width=2cm}] (Tr1) [below of=Pi] {};
  \node[main node,style={draw,dashed,shape border uses incircle,isosceles triangle,shape border rotate=90,yshift=-0.25cm,scale=1.8,minimum width=2cm}] (Tr2) [below of=Pj] {};
  \node[main node,style={draw,dashed,shape border uses incircle,isosceles triangle,shape border rotate=90,yshift=-0.25cm,scale=1.8,minimum width=2cm}] (Tr3) [below of=Pk] {};
  \node[main node] (Ty) [below of=Tr2,yshift=2cm,xshift=0.5cm,scale=0.85] {$T_y$};
  \node[main node] (Tx) [below of=Tr3,yshift=2cm,xshift=-0.5cm,scale=0.85] {$T_x$};

  \draw[-,very thick,decorate,decoration={brace,amplitude=15pt,raise=0.75cm}] (P1.north east) -- (Pi.north east) node [black,midway,yshift=1.0cm,xshift=1.0cm] {\LARGE$\pi_{pre}$};

  \path[every node/.style={font=\sffamily\small,fill=white,inner sep=1pt}]
    (P0) edge [bend left=0,line width=1.0mm] node[right=1mm] {} (P1)
    (P1) edge [bend left=0] node[right=1mm] {} (Pk)
    (Pbr) edge [bend left=0] node[right=1mm] {} (Pj)
    (Pbr) edge [bend left=0,line width=1.0mm] node[right=1mm] {} (Pi);

  \draw[line width=1.0mm,decorate,decoration={snake,amplitude=.8mm,segment length=8mm,post length=2mm}] (P1) -- (Pbr);
  \draw[line width=1.0mm,decorate,decoration={snake,amplitude=.8mm,segment length=8mm,post length=2mm}] (Tr1.north) -- ([xshift=0.7cm]Tr1.south);
\end{tikzpicture}
}
\caption{Execution tree of a program containing target locations $T_x$
  and $T_y$. The lookahead analysis analyzes a path $\pi$ (bold) to
  identify a prefix $\pi_{pre}$ such that no suffix paths reach a
  target location.}
\label{fig:diagram}
\vspace{-1em}
\end{figure}

\textbf{Our approach.} In this paper, we present a technique that
\emph{semantically} guides greybox fuzzing toward \emph{target
  locations}, for instance, locations reported by another
analysis or located in recently modified parts of the program. This is
achieved with an \emph{online} static analysis.
In particular, the fuzzer invokes this online analysis right before
adding a new input to its test suite. For the program path $\pi$ that
the new input explores (see bold path in Fig.~\ref{fig:diagram}), the
goal of the analysis is to determine a path prefix $\pi_{pre}$ for
which all suffix paths are unable to reach a target location (e.g.,
$T_x$ and $T_y$ in Fig.~\ref{fig:diagram}). This additional
information allows the fuzzer to allocate its resources more
strategically such that more effort is spent on exercising program
paths that might reach the target locations, thereby enabling
\emph{targeted fuzzing}. More precisely, this information feeds into a
specialized power schedule of the fuzzer that determines how often to
fuzz an input from the test suite.

We refer to our online static analysis as a \emph{lookahead analysis}
since, given a path prefix $\pi_{pre}$, it looks for reachable target
locations along all suffix paths (sub-tree rooted at $P_i$ in
Fig.~\ref{fig:diagram}). We call the last program location of prefix
$\pi_{pre}$ a \emph{split point} ($P_i$ in
Fig.~\ref{fig:diagram}). Unlike a traditional static analysis, the
lookahead analysis does not consider all possible execution states at
the split point when analyzing all suffix paths---only the ones that
are feasible along $\pi_{pre}$. In other words, the lookahead analysis
combines the precision of a path-sensitive analysis along a feasible
path prefix with the scalability of a path-insensitive suffix
analysis. Intuitively, for a given path $\pi$, the precision of the
lookahead analysis is determined by the number of suffix paths that
are proved not to reach any target locations. Therefore, to optimize
precision, the analysis tries to identify the \emph{first} split point
($P_i$ in Fig.~\ref{fig:diagram}) along $\pi$ such that all targets
are unreachable.
Note that the lookahead analysis may consider any program location along
$\pi$ as a split point.

When combining greybox fuzzing with an online lookahead analysis, we faced four main
challenges, which we address in this paper. In particular, we provide answers to the
following questions: (1)~How can the lookahead analysis effectively communicate its
results to the fuzzer?  (2)~How lightweight can the analysis be to improve the
effectiveness of the fuzzer in reaching target locations without having a negative impact
on its performance? (3)~How can the analysis be invoked from a certain split point along a
path? (4)~What are suitable split points for invoking the analysis to check all suffix paths?

Our implementation uses \harvey, a state-of-the-art, industrial
greybox fuzzer for Ethereum smart contracts, which are programs
managing crypto-currency accounts on a blockchain. We extended \harvey
to incorporate \bran, a new static-analysis framework for smart
contracts.
A main reason for targeting the domain of smart contracts is that
adding code instrumentation to contracts changes their semantics, and
all existing techniques that use an offline static analysis require
instrumentation of the program under test.
Our experiments on 27 benchmarks show that targeted fuzzing
significantly outperforms standard greybox fuzzing for reaching 83\%
of the challenging target locations (up to 14x of median speed-up).

\textbf{Contributions.} We make the following contributions:
\begin{itemize}
\item We introduce a greybox-fuzzing algorithm that uses a
  lightweight, online static analysis and a specialized power schedule
  to guide the exploration toward target locations.

\item We implement this fuzzing algorithm by extending the \harvey
  greybox fuzzer with \bran, a static analysis for smart contracts.

\item We evaluate our technique on 27 real-world benchmarks and
  demonstrate that our lookahead analysis and power schedule
  significantly increase the effectiveness of greybox fuzzing in
  reaching target locations.
\end{itemize}

\textbf{Outline.} The next section provides background on greybox
fuzzing and smart contracts. In Sect.~\ref{sect:overview}, we give an
overview of our technique through an
example. Sect.~\ref{sect:technique} explains the technical details,
and Sect.~\ref{sect:implementation} describes our implementation.  We
present our experimental evaluation in Sect.~\ref{sect:experiments},
discuss related work in Sect.~\ref{sect:relatedWork}, and conclude in
Sect.~\ref{sect:conclusion}.

\section{Background}
\label{sect:background}

In this section, we review background on greybox fuzzing and smart
contracts.

\subsection{Greybox Fuzzing}
\label{subsect:fuzzing}

Greybox fuzzing~\cite{AFL,LibFuzzer} is a practical test-generation
technique that has been shown to be very effective in detecting bugs
and security vulnerabilities (e.g.,
\cite{AFL-Bugs}). Alg.~\ref{alg:greyboxFuzzingWithLookaheadAnalysis}
shows exactly how it works. (The grey boxes should be ignored.)

A greybox fuzzer takes as input the program under test $\mathit{prog}$
and a set of seed inputs $S$. The fuzzer runs the program with the
seeds (line~1) and associates each input with the unique identifier of
the path it exercises, or $\mathit{PID}$. The $\mathit{PIDs}$ data
structure, therefore, represents a map from a $\mathit{PID}$ to the
corresponding input. Note that a path identifier is computed with
lightweight runtime monitoring that allows the fuzzer to distinguish
different program paths.

Next, the fuzzer selects an input from $\mathit{PIDs}$ for mutation
(line~3), which is typically performed randomly. This input is
assigned an ``energy'' value, which indicates how long it should be
fuzzed (line~5). The input is then mutated (line~8), and the program
is run again with this new input (line~9). If the new input exercises
a path that has not been seen before, it is added to $\mathit{PIDs}$
with the corresponding path identifier (lines~10, 12).

This process terminates when a bound is reached, such as a timeout or
a number of generated inputs (line~2). When that happens, the fuzzer
returns a test suite comprising all inputs in $\mathit{PIDs}$, each
exercising a different path in the program.

\subsection{Smart Contracts}

Ethereum~\cite{EthereumWhitePaper} is one of the most well known
blockchain-based~\cite{BlockchainBlueprint,BlockchainTechnology}
computing platforms. Like a bank, Ethereum supports accounts that
store a balance (in digital assets) and are owned by a user. More
specifically, there is support for two account types, namely user and
contract accounts.

Contract accounts are not managed by a user, but instead by a
program. The program associated with a certain contract account
describes an agreement between the account and any users that interact
with it. For example, such a program could encode the rules of a
gambling game. To store information, such as bets from various users,
a contract account also comes with persistent state that the program
may access and modify.

A contract account together with its managing program and persistent
state is called a \emph{smart contract}. However, the term may also
refer to the code alone. Ethereum smart contracts can be developed in
several high-level languages, such as Solidity and Vyper, which
compile to Ethereum Virtual Machine (EVM)~\cite{EthereumYellowPaper}
bytecode.

Users interact with a smart contract, for instance to place a bet, by
issuing a transaction with the contract. The transaction simply calls
one of the contract functions, but in order to be carried out, users
need to provide a fee. This fee is called \emph{gas} and is
approximately proportional to how much code needs to run. Any
transaction that runs out of gas is aborted.

\section{Overview}
\label{sect:overview}

We now give an overview of our approach through the example
of Fig.~\ref{fig:example}.

\textbf{Example.} The figure shows a constructed function \code{Bar},
which is written in Solidity and contained in a smart contract. (The
comments should be ignored for now.)
There are three assertions in this function, on
lines~\ref{line:assert1}, \ref{line:assert2}, and
\ref{line:assert3}. A compiler will typically introduce a conditional
jump for each assertion, where one branch leads to a location that
fails. Let us assume that we select the failing locations
($t_{\ref{line:assert1}}$, $t_{\ref{line:assert2}}$ , and
$t_{\ref{line:assert3}}$) of the three assertions as our target
locations.
Note that any target locations could be (automatically) selected based
on various strategies, e.g., recently modified code, assertions, etc.
Out of the above locations, $t_{\ref{line:assert1}}$ and
$t_{\ref{line:assert2}}$ are unreachable, whereas
$t_{\ref{line:assert3}}$ is reachable when input parameter \code{a}
has value 42.

Generating a test input that reaches location $t_{\ref{line:assert3}}$
is difficult for a greybox fuzzer for two reasons. First, the
probability of generating value 42 for parameter \code{a} is tiny,
namely $1$ out of $2^{256}$. This means that, for the fuzzer to
increase the chances of reaching $t_{\ref{line:assert3}}$, it would
need to fuzz certain ``promising'' inputs with a large amount of
energy. However, standard greybox fuzzers are agnostic to what
constitutes a promising input that is more likely to reach a target
location when mutated.

Second, there are more than 100'000 program paths in function
\code{Bar}.
In fact, the then-branch of the first if-statement
(line~\ref{line:if1}) contains two input-dependent loops
(lines~\ref{line:loop1} and \ref{line:loop2}), whose number of
iterations depends on parameters \code{w} and \code{z}, respectively.
Recall that a greybox fuzzer generates new inputs by mutating existing
ones from the test suite. Therefore, the larger the size of the test
suite, the larger the space of possible mutations, and the lower the
chances of generating an input that reaches the target location.

\begin{figure}[t]
\begin{lstlisting}[style=solidity]
function Bar(uint256 w, uint256 x, uint256 y,
      uint256 z, uint256 a) returns (uint256)
{
  uint256 ret = 0;
  if (x % 2 == 0) {   // ¤\color{purple}\textbf{if}¤ (x % 1000 != 42) {  ¤\label{line:if1}¤
    ret = 256;
    if (y % 2 == 0) { ¤\label{line:if2}¤
      ret = 257;
    } ¤\label{line:end-if2}¤
    w = w % ret;
    while (w != 0) { ¤\label{line:loop1}¤
      w--;
    }
    assert(w == 0);   // ¤\color{purple}\textit{drop this line}¤ ¤\label{line:assert1}¤
    z = z % ret; ¤\label{line:z}¤
    while (ret != z) { ¤\label{line:loop2}¤
      z++;
    }
    assert(ret == z); // ¤\color{purple}\textbf{assert}¤(x != 42 - w*z); ¤\label{line:assert2}¤
  } else {
    ret = 3*a*a + 7*a + 101; ¤\label{line:rare-loc}¤
    assert(ret != 5687); ¤\label{line:assert3}¤
  }
  return ret;
}
\end{lstlisting}
\caption{The running example.}
\label{fig:example}
\end{figure}

\textbf{Existing work.}
As discussed earlier, there is existing work that leverages the
results of an offline static analysis to guide automatic test
generation toward unverified executions (e.g.,
\cite{CsallnerSmaragdakis2005,GeTaneja2011,CzechJakobs2015,ChristakisMueller2016,FerlesWuestholz2017}). To
apply such a technique on the example of Fig.~\ref{fig:example}, let
us assume a very lightweight static analysis that is based on abstract
interpretation~\cite{CousotCousot1977,CousotCousot1979} and uses the
simple constant-propagation domain~\cite{Kildall1973}. Note that, for
each program variable, the constant-propagation domain can only infer
a single constant value.
When run offline, this analysis is able to prove that target
location $t_{\ref{line:assert1}}$ is unreachable. This is because,
after the loop on line~\ref{line:loop1}, the analysis assumes the
negation of the loop condition (that is, \code{w == 0}), which is
equivalent to the asserted condition.

However, the analysis cannot prove that location
$t_{\ref{line:assert2}}$ is also unreachable. This is because, after
the if-statement on line~\ref{line:if2}, variable \code{ret} has
abstract value $\top$. In other words, the analysis finds \code{ret}
to be unconstrained since the constant-propagation domain is not able
to express that its value is either 256 or 257. Given that \code{ret}
is $\top$, \code{z} also becomes $\top$ (line~\ref{line:z}). It is,
therefore, not possible for the analysis to determine whether these
two variables always have the same value on line~\ref{line:assert2}
and verify the assertion.
As a result, automatic test generation needs to explore function
\code{Bar} as if no static analysis had previously run. To check
whether the assertion on line~\ref{line:assert2} always holds, a
testing tool would have to generate inputs for all paths leading to
it, thus including each iteration of the loop on
line~\ref{line:loop1}.

On the other hand, an existing technique for directed greybox
fuzzing~\cite{BoehmePham2017} performs lightweight instrumentation of
the program under test to extract a distance metric for each input,
which is then used as feedback for the fuzzer. So, the instrumentation
encodes a static metric that measures the distance between the
instrumented and the target locations in the control-flow graph. In
our example, such metrics are less effective since all instructions
are relatively close to the target locations, and the control-flow
graph alone is not precise enough to determine more semantic
reachability conditions. In addition, when directly fuzzing bytecode
or assembly, a control-flow graph might not be easily recoverable, for
instance due to indirect jumps.

\textbf{Lookahead analysis.}
In contrast, our approach alleviates the imprecision of a static
analysis by running it online and does not require a control-flow
graph.
Our greybox fuzzer invokes the lookahead analysis for each input that
is added to the test suite.
Starting from split points (e.g., $P_1$, $P_{i-1}$, and $P_i$ in
Fig.~\ref{fig:diagram}) along an explored program path, the analysis
computes a path prefix ($\pi_{pre}$) for which all suffix paths do not
reach any target location (e.g., $T_x$ and $T_y$).
We refer to such a path prefix as a \emph{no-target-ahead prefix}
(see Def.~\ref{def:no-target-head-prefix} for more details). As we
explain below, the lookahead analysis aims to identify short
no-target-ahead prefixes.

As an example, let us consider the constant-propagation analysis and
an input for function \code{Bar} with an even value for \code{x} (thus
making execution take the then-branch of the first if-statement on
line~\ref{line:if1}). Along the path exercised by this input, the
analysis fails to show that both target locations
$t_{\ref{line:assert1}}$ and $t_{\ref{line:assert2}}$ are unreachable
for the suffix paths starting from line~\ref{line:if2}. In fact, the
analysis is as imprecise as when run offline on the entire
function. However, it does verify the unreachability of the target
locations for all suffix paths from line~\ref{line:end-if2} by
propagating forward the constant value of variable \code{ret} (either
256 or 257, depending on the value of \code{y}). Out of the many paths
with an even value for \code{x}, the two no-target-ahead prefixes
until line~\ref{line:end-if2} (through the then- and else-branches of
the if-statement on line~\ref{line:if2}) are actually the shortest
ones for which the lookahead analysis proves that target locations
$t_{\ref{line:assert1}}$ and $t_{\ref{line:assert2}}$ are unreachable.

\textbf{Power schedule.}
The no-target-ahead prefixes computed by the lookahead analysis are
used to control the fuzzer's power schedule~\cite{BoehmePham2016},
which assigns more energy to certain inputs according to two criteria.

First, it assigns more energy to inputs that exercise a rare (i.e.,
rarely explored) no-target-ahead prefix. The intuition is to fuzz
these inputs more in order to increase the chances of flipping a
branch along the rare prefix, and thereby, reaching a target
location. Note that flipping a branch in a suffix path can never lead
to a target location. For this reason, our power schedule no longer
distinguishes inputs based on the program path they exercise, but
rather based on their no-target-ahead prefix. To maximize the chances
of discovering a target location with fuzzing, the lookahead analysis
tries to identify the shortest no-target-ahead prefixes, which are
shared by the most suffix paths.

For the example of Fig.~\ref{fig:example}, consider the two
no-target-ahead prefixes (until line~\ref{line:end-if2}) that we
discussed above. Consider also the no-target-ahead prefix until the
successful branch of the assertion on line~\ref{line:assert3}. The
inputs that exercise these prefixes are dynamically assigned roughly
the same energy by our schedule---if one of them is exercised more
rarely than the others, it is given more energy. This makes reaching
target location $t_{\ref{line:assert3}}$ significantly more likely
than with standard power schedules based on path identifiers, which
assign roughly the same energy to each input exercising one of the
thousands of paths in \code{Bar}.

Second, our power schedule also assigns more energy to inputs
exercising rare split points in a no-target-ahead prefix, similarly to
how existing work assigns more energy to rare
branches~\cite{LemieuxSen2018}. The intuition is the following. Any
newly discovered no-target-ahead prefix is by definition rare---it has
not been fuzzed before. Since it is rare, the power schedule will
assign more energy to it, as discussed above. However, there are
programs where new no-target-ahead prefixes can be easily discovered,
for instance due to an input-dependent loop. In such cases, a power
schedule only focusing on rare prefixes would prioritize these new
prefixes at the expense of older ones that explore rare program
locations, such as split points.  For this reason, when a split point
in a no-target-ahead prefix becomes rare, the power schedule tries to
explore it more often.

As an example, consider the code in Fig.~\ref{fig:example} while
taking the comments into account, that is, replace
lines~\ref{line:if1} and \ref{line:assert2} with the comments and drop
line~\ref{line:assert1}. The assertion on line~\ref{line:assert2}
holds, but the constant-propagation analysis is too weak to prove it.
As a result, for any path through this assertion, its no-target-ahead
prefix has to include line~\ref{line:assert2}. However, new
no-target-ahead prefixes are very easily discovered; for instance, by
exploring a different number of iterations in any of the two loops.
So, even if at some point the fuzzer discovers the path that
successfully exercises the assertion on line~\ref{line:assert3}, its
no-target-ahead prefix will quickly become less rare than any new
prefixes going through the loops.
The corresponding input will, therefore, be fuzzed less often even
though it is very close to revealing the assertion violation.
By prioritizing rare split points, for instance
line~\ref{line:rare-loc}, our power schedule will assign more energy
to that input.
This increases the chances of mutating the value of \code{a} to be 42
and reaching target $t_{\ref{line:assert3}}$.

Both of these criteria effectively guide the fuzzer toward the target
locations. For Fig.~\ref{fig:example}, our technique generates a test
that reaches $t_{\ref{line:assert3}}$ in 27s on average (between 13
and 48s in 5 runs). Standard greybox fuzzing does not reach
$t_{\ref{line:assert3}}$ in 4 out of 5 runs, with a timeout of
300s. The target location is reached in 113s during a fifth run, so in
263s on average. For this example, our technique achieves at least a
10x speed-up.

\textbf{Why smart contracts.}
While our approach could be applied to regular programs, it is
particularly useful in the context of smart contracts. One reason is
that, in this setting, combining an offline static analysis with test
generation using code instrumentation would change the program
semantics. Recall that a transaction with a smart contract is carried
out when users provide enough gas, which is roughly proportional to
how much code is run. Since instrumentation consumes gas at execution
time, it could cause a testing tool to report spurious out-of-gas
errors. Another reason is that most deployed smart contracts are only
available as bytecode, and recovering the control-flow graph from the
bytecode is challenging.

\section{Technique}
\label{sect:technique}



In this section, we describe our technique in detail by first formally
defining a lookahead analysis
(Sect.~\ref{subsect:lookahead-analysis}). We then discuss how to
integrate such an analysis with greybox fuzzing to enable a more
targeted exploration of the search space
(Sect.~\ref{subsect:lookahead-fuzzing}). Lastly, we present a concrete
algorithm for a lookahead analysis based on abstract interpretation.

\subsection{Lookahead Analysis}
\label{subsect:lookahead-analysis}

Let us first define a prefix and a no-target-ahead prefix of a given
path.

\begin{definition}[\bf Prefix] \label{def:prefix}
  Given a program $P$ and a path $\pi$ in $P$, we say that
  $\pi_{\mathit{pre}}$ is a \emph{prefix} of $\pi$ iff there exists a
  suffix $\rho$ such that $\pi = \mathit{concat}(\pi_{\mathit{pre}},
  \rho)$.
\end{definition}

Note that, in the above definition, $\rho$ may be empty, in which case
$\pi = \pi_{\mathit{pre}}$.

\begin{definition}[\bf No-target-ahead prefix] \label{def:no-target-head-prefix}
  Given a program $P$, target locations $T$, and a prefix
  $\pi_{\mathit{pre}}$ of a path in $P$, we say that
  $\pi_{\mathit{pre}}$ is a \emph{no-target-ahead prefix} iff the
  suffix $\rho$ of every path $\pi =
  \mathit{concat}(\pi_{\mathit{pre}}, \rho)$ in $P$ does not contain a
  target location $\tau \in T$.
\end{definition}

Note that any path $\pi$ in a program $P$ is trivially a
no-target-ahead prefix since there cannot be any target locations
after reaching the end of its execution.

For a given no-target-ahead prefix, the analysis computes a
\emph{lookahead identifier} ($\mathit{LID}$) that will later be used
to guide the greybox fuzzer.

\begin{definition}[\bf Lookahead identifier] \label{def:lid}
  Given a no-target-ahead prefix $\pi_{\mathit{pre}}$, the
  \emph{lookahead identifier} $\lambda$ is a cryptographic hash
  $\mathit{hash}(\pi_{\mathit{pre}})$.
\end{definition}

The above definition ensures that it is very unlikely that two
different no-target-ahead prefixes map to the same lookahead
identifier.

Unlike a path identifier ($\mathit{PID}$) in standard greybox fuzzing,
which is computed purely syntactically, a $\mathit{LID}$ captures a
no-target-ahead prefix, which is computed by semantically analyzing a
program path.
As a result, two program paths with different $\mathit{PIDs}$ may
share the same $\mathit{LID}$. In other words, lookahead identifiers
define equivalence classes of paths that share the same
no-target-ahead prefix.

\begin{definition}[\bf Lookahead analysis] \label{def:analysis}
  Given a program $P$, an input $I$, and a set of target locations
  $T$, a \emph{lookahead analysis} computes a lookahead identifier
  $\lambda$ for the corresponding no-target-ahead prefix
  $\pi_{\mathit{pre}}$ (of path $\pi$ exercised by input $I$) and a
  set of split points $\mathit{SPs}$ along $\pi_{\mathit{pre}}$.
\end{definition}

Any analysis that satisfies the above definition is a sound lookahead
analysis. For instance, one that simply returns the hash of path $\pi$
exercised by input $I$ and all locations along $\pi$ is trivially
sound. For a given input, the precision of the analysis is determined
by the length of the no-target-ahead prefix, and thereby, the number
of suffix paths that are proved not to contain any target locations.
In other words, the shorter the no-target-ahead prefix for a given
input, the more precise the lookahead analysis.

\subsection{Fuzzing with Lookahead Analysis}
\label{subsect:lookahead-fuzzing}

The integration of greybox fuzzing with a lookahead analysis builds on
the following core idea. For each input in the test suite, the
lookahead analysis determines a set of split points, that is, program
locations along the explored path. It then computes a no-target-ahead
prefix, which spans until one of these split points and is identified
by a lookahead identifier. The fuzzer uses the rarity of the lookahead
identifier as well as of the split points that are located along the
no-target-ahead prefix to assign energy to the corresponding input.

The grey boxes in Alg.~\ref{alg:greyboxFuzzingWithLookaheadAnalysis}
highlight the key extensions we made to standard greybox fuzzing. For
one, our algorithm invokes the lookahead analysis on line~11. This is
done for every new input that is added to the test suite and computes
the $\mathit{LID}$ of the no-target-ahead prefix as well as the split
points $\mathit{SPs}$ along the prefix. Both are stored in the
$\mathit{PIDs}$ data structure for efficient lookups (e.g.,
when assigning energy).

\begin{algorithm}[t]
  \caption{\textbf{Greybox fuzzing with lookahead analysis.}}
  \label{alg:greyboxFuzzingWithLookaheadAnalysis}
  \hspace{-0em}\textbf{Input:} Program $\mathit{prog}$, Seeds $S${\btHL[fill=light-gray], Target locations $T$}
  \begin{algorithmic}[1]
    \small
      \Let{$\mathit{PIDs}$}{\Fcall{RunSeeds}$(S, \mathit{prog})$}
      \While{$\neg$\Fcall{Interrupted}()}
        \Let{$\mathit{input}$}{\Fcall{PickInput}$(\mathit{PIDs})$}
        \Let{$\mathit{energy}$}{0}
        \Let{$\mathit{maxEnergy}$}{\Fcall{AssignEnergy}$(\mathit{input})$}
        \LetHL{$\mathit{maxEnergy}$}{\Fcall{LookaheadAssignEnergy}$(\mathit{input})$}
        \While{$\mathit{energy} < \mathit{maxEnergy}$}
          \Let{$\mathit{input'}$}{\Fcall{FuzzInput}$(\mathit{input})$}
          \Let{$\mathit{PID'}$}{\Fcall{Run}$(\mathit{input'}, \mathit{prog})$}
          \If{\Fcall{IsNew}($\mathit{PID'}, \mathit{PIDs}$)}
            \LetHL{$\mathit{LID}, \mathit{SPs}$}{\Fcall{LookaheadAnalyze}$(\mathit{prog}, \mathit{input'}, T)$}
            \Let{$\mathit{PIDs}$}{\Fcall{Add}$(\mathit{PID'}, \mathit{input'}, {\btHL[fill=light-gray] $\mathit{LID}, \mathit{SPs},\,$} \mathit{PIDs})$}
          \EndIf
          \Let{$\mathit{energy}$}{$\mathit{energy} + 1$}
        \EndWhile
      \EndWhile
  \end{algorithmic}
  \hspace{-0em}\textbf{Output:} Test suite \textsc{Inputs}$(\mathit{PIDs})$
\end{algorithm}

We also replace the existing power schedule on line~5 with a
specialized one given by $\Fcall{LookaheadAssignEnergy}$ (line~6). As
discussed in Sect.~\ref{sect:overview}, our power schedule assigns
more energy to inputs that exercise either a \emph{rare
  $\mathit{LID}$} or a \emph{rare split point} along a no-target-ahead
prefix. We define the new power schedule in the following.

\begin{definition}[\bf Rare \textit{LID}] \label{def:rare_LID}
  Given a test suite with $\mathit{LIDs}$ $\Lambda$, a $\mathit{LID}$
  $\lambda$ is rare iff
  \[\mathit{fuzz}(\lambda) < \mathit{rarity\_cutoff},\]
  where $\mathit{fuzz(\lambda)}$ measures the number of fuzzed inputs
  that exercised $\lambda$ so far and $\mathit{rarity\_cutoff} = 2^i$
  such that
  \[2^{i-1} < \min_{\lambda' \in \Lambda}\mathit{fuzz}(\lambda') \leq 2^i.\]
\end{definition}

For example, if the $\mathit{LID}$ with the fewest fuzzed inputs has
been explored 42 times, then any $\mathit{LID}$ that has been explored
less than $2^6$ times is rare.

The above definition is inspired by an existing power schedule for
targeting rare branches~\cite{LemieuxSen2018} that introduced such a
dynamically adjusted $\mathit{rarity\_cutoff}$. Their experience shows
that this metric performs better than simply considering the $n$
$\mathit{LIDs}$ with the lowest number of fuzzed inputs as rare.

\begin{definition}[\bf Rare split point] \label{def:rare_split_point}
  Given a test suite with split points $\mathit{SPs}$ along the
  no-target-ahead prefixes, a split point $p$ is rare iff
  \[\mathit{fuzz}(p) < \mathit{rarity\_cutoff},\]
  where $\mathit{fuzz(p)}$ measures the number of fuzzed
  inputs that exercised $p$ so far and $\mathit{rarity\_cutoff} = 2^i$ such that
  \[2^{i-1} < \min_{p' \in \mathit{SPs}}\mathit{fuzz}(p') \leq 2^i.\]
\end{definition}

\textbf{Power schedule.} Our power schedule is defined as follows for
an input $I$ with $\mathit{LID}$ $\lambda$ and split points
$\mathit{SPs}$ along the no-target-ahead prefix:
 \[
  \begin{cases}
    \min(2^{selected(I)}, K), & \text{if }\lambda \text{ is rare} \vee \exists p \in \mathit{SPs} \cdot p \text{ is rare}\\
    1, & \text{otherwise}.
  \end{cases}
  \]
In the above definition, $\mathit{selected(I)}$ denotes the number of
times that $I$ was selected for fuzzing (line~3 in
Alg.~\ref{alg:greyboxFuzzingWithLookaheadAnalysis}), and $K$ is a
constant (1024 in our implementation). Intuitively, our power schedule
assigns little energy to inputs whose $\mathit{LID}$ is not rare and
whose no-target-ahead prefix does not contain any rare split
points. Otherwise, it assigns much more energy, the amount of which
depends on how often the input has been selected for fuzzing
before. The energy grows exponentially up to some bound $K$, similarly
to the cut-off-exponential schedule in AFLFast~\cite{BoehmePham2016}.

\subsection{Lookahead Algorithm}
\label{subsect:lookahead-algorithm}

\begin{algorithm}[t]
  \caption{\textbf{Lookahead algorithm.}}
  \label{alg:LookaheadAnalysis}
  \hspace{-0em}\textbf{Input:} Program $\mathit{prog}$, Input $\mathit{input}$, Target locations $T$
  \begin{algorithmic}[1]
    \small
      \Let{$\pi$}{\Fcall{Run}$(\mathit{input}, \mathit{prog})$}
      \Let{$\mathit{i}$}{$0$}
      \Let{$\mathit{SPs}$}{$\emptyset$}
      \While{$\mathit{i} < |\pi|$}
        \If{\Fcall{IsSplitPoint}($\mathit{i}, \pi)$}
          \Let{$\pi_{\mathit{pre}}$}{$\pi[0..i+1]$}
          \Let{$\mathit{SPs}$}{$\mathit{SPs} \cup \{\pi[i]\}$}
          \Let{$\phi, \mathit{loc}$}{\Fcall{PrefixInference}$(\pi_{\mathit{pre}})$}
          \If{\Fcall{AreTargetsUnreachable}$(\mathit{prog}, \mathit{loc}, \phi, T$)}
            \LRet{\Fcall{ComputeHash}$(\pi_{\mathit{pre}})$, $\mathit{SPs}$}
          \EndIf
        \EndIf
        \Let{$\mathit{i}$}{$\mathit{i} + 1$}
      \EndWhile
      \LRet{\Fcall{ComputeHash}$(\pi)$, $\mathit{SPs}$}
  \end{algorithmic}
  \hspace{-0em}\textbf{Output:} Lookahead identifier $\lambda$, Split points $\mathit{SPs}$
\end{algorithm}

Alg.~\ref{alg:LookaheadAnalysis} shows the algorithm for the lookahead
analysis, which is implemented in function \textsc{LookaheadAnalyze}
from Alg.~\ref{alg:greyboxFuzzingWithLookaheadAnalysis} and uses
abstract interpretation~\cite{CousotCousot1977,CousotCousot1979}.

First, the lookahead analysis executes the program input concretely to
collect the exercised path $\pi$ (line~1 in
Alg.~\ref{alg:LookaheadAnalysis}). Given path $\pi$, it searches for
the shortest no-target-ahead prefix $\pi_{\mathit{pre}}$ by iterating
over possible split points $p$ (lines~4--11). Let us explain these
lines in detail.

On line~5, the algorithm calls a predicate $\Fcall{IsSplitPoint}$,
which is parametric in which locations constitute split points. All
locations along $\pi$ could be split points, but to narrow down the
search, the implementation may consider only a subset of them, for
instance, at conditional jumps.

At each split point, the analysis performs two separate steps:
(1)~\emph{prefix inference} and (2)~\emph{suffix checking}. The prefix
inference (line~8) statically analyzes the prefix $\pi_{\mathit{pre}}$
using abstract interpretation to infer its postcondition $\phi$. This
step essentially executes the prefix in the abstract for all possible
inputs that exercise this path.

Given condition $\phi$, the analysis then performs the suffix checking
to determine if all target locations are unreachable (line~9). This
analysis performs standard, forward abstract interpretation by
computing a fixed-point. If all target locations are unreachable, the
analysis terminates and returns a non-empty $\mathit{LID}$ by
computing a hash over the program locations along the path prefix
$\pi_{\mathit{pre}}$ (line~10). This ensures that the analysis returns
as soon as it reaches the first split point for which all targets are
unreachable. In addition, it returns the set of all split points along
prefix $\pi_{\mathit{pre}}$.

Even though off-the-shelf abstract interpreters are not designed to
perform prefix inference and suffix checking, it is relatively
straightforward to extend them. Essentially, when invoking a standard
abstract interpreter on a program, the path prefix is always empty,
whereas our lookahead analysis is partially path-sensitive (i.e., for
the prefix, but not the suffix). Due to this partial path-sensitivity,
even an inexpensive abstract domain (e.g., constant propagation or
intervals) might be able to prove unreachability of a certain target
location, which would otherwise require a more precise domain (for an
empty prefix).

\textbf{Split points.} In practice, it is important to choose split
points with care since too many split points will have a negative
impact on the performance of the lookahead analysis. In our
implementation, we only consider split points when entering a basic
block for the first time along a given path. The intuition is that the
lookahead analysis should run every time ``new code'' is
discovered. Our experiments show that this design decision results in
negligible overhead.

\textbf{Calls.} To keep the lookahead analysis lightweight, the
suffix-checking step is modular. More specifically, any calls to other
contracts are conservatively treated as potentially leading to target
locations. (Note that inter-contract calls are used very sparingly in
smart contracts and that intra-contract calls are simply jumps.) In
contrast, during the prefix-inference step, we compute a summary
$\mathit{LID}$ for the callee context by recursively invoking the
lookahead algorithm on the callee. This requires separating the parts
of path $\pi$ (from Alg.~\ref{alg:LookaheadAnalysis}) that belong to
the caller and the callee. It is also necessary to conservatively
model the effect of a call on the caller context (e.g., by havocking
return values).

\section{Implementation}
\label{sect:implementation}

Our implementation extends \harvey~\cite{WuestholzChristakis2018Harvey,WuestholzChristakis2019Harvey},
an existing greybox fuzzer for Ethereum smart contracts. It is actively used at
ConsenSys Diligence, one of the largest blockchain-security consulting companies, and is
one of the tools that power the MythX analysis platform.
For our purposes, we integrated \harvey with \bran, our new
abstract-interpretation framework for EVM bytecode, which is open
source\footnote{\url{https://github.com/Practical-Formal-Methods/bran}}.

\bran is designed to be scalable by performing a very lightweight,
modular analysis that checks functional-correctness properties. Unlike
other static analyzers for EVM bytecode (e.g.,
Securify~\cite{TsankovDan2018} and MadMax~\cite{GrechKong2018}), \bran
runs directly on the bytecode without having to reconstruct the
control-flow graph or decompile to an intermediate language.
\bran is equipped with a constant-propagation
domain~\cite{Kildall1973}, which is commonly used in compiler
optimizations. It handles all opcodes and integrates the go-ethereum
virtual machine to concretely execute any opcodes with all-constant
arguments.

\textbf{Prefix length.} During our preliminary experiments with the
integration of \harvey and \bran, we observed that the prefix length
may become quite large, for instance in the presence of
input-dependent loops. However, the running time of the lookahead
analysis is proportional to the prefix length, and our goal is to keep
the analysis as lightweight as possible. For this reason, our
implementation ignores any split points after the first 8'192 bytecode
locations of the prefix. Note that this design decision does not
affect the soundness of the lookahead analysis; it only reduces the
search space of prefixes and might result in considering the entire
path as the no-target-ahead prefix.

\section{Experimental Evaluation}
\label{sect:experiments}

We now evaluate our technique on real-world Ethereum smart
contracts. First, we discuss the benchmark selection
(Sect.~\ref{subsect:benchmarks}) and describe our experimental setup
(Sect.~\ref{subsect:setup}). We then evaluate the effectiveness of the
static lookahead analysis in greybox fuzzing
(Sect.~\ref{subsect:results}) and identify potential threats to the
validity of our experiments (Sect.~\ref{subsect:threats}).

\subsection{Benchmark Selection}
\label{subsect:benchmarks}

We evaluated our technique on a total of 27 smart contracts, which
originate from 17 GitHub repositories. Tab.~\ref{tab:benchmarks} gives
an overview. The first column lists a benchmark identifier for each
smart contract under test, while the second and last columns provide
the name and description of the containing project. Note that a
repository may contain more than one contract, for instance including
libraries; from each repository, we selected one or more contracts for
our evaluation. The third and fourth columns of the table show the
number of public functions and lines of Solidity code in each
benchmark. (We provide links to all repositories as well as the
changesets used for our experiments in the appendix.)

It is important to note that the majority of smart contracts are under
1'000 lines of code. Still, contracts of this size are complex
programs, and each of them might take several weeks to audit. However,
as it becomes clear from the example of Fig.~\ref{fig:example}, code
size is not necessarily proportional to the number of feasible program
paths or the difficulty to reach a particular target location with
greybox fuzzing.

The repositories were selected with the goal of ensuring a diverse set
of benchmarks. In particular, they include popular projects, such as
the ENS domain name auction, the ConsenSys multisig wallet, and the
MicroRaiden payment service. In addition to being widely known in the
Ethereum community, these projects are highly starred on GitHub (4'857
stars in total on 2019-05-07, median 132), have been independently
audited, and regularly transfer large amounts of assets.
Moreover, our selection includes contracts from various application
domains (like auctions, wallets, and tokens), attacked contracts
(namely, The DAO and Parity wallet) as well as contracts submitted to
the first Underhanded Solidity Coding Contest
(USCC)~\cite{USCC}. Entries in this contest aim to conceal subtle
vulnerabilities.

For selecting these repositories, we followed guidelines on how to
evaluate fuzzers~\cite{KleesRuef2018}. We do not randomly collect
smart contracts from the Ethereum blockchain since this would likely
contaminate our benchmarks with duplicates or bad-quality
contracts---that is, contracts without users, assets, or dependencies,
for instance, on libraries or other contracts.

\begin{table}[t!]
\centering
\scalebox{0.95}{
\begin{tabular}{r|l|r|r|l}
\multicolumn{1}{c|}{\textbf{BIDs}} & \multicolumn{1}{c|}{\textbf{Name}} & \multicolumn{1}{c|}{\textbf{Functions}} & \multicolumn{1}{c|}{\textbf{LoSC}} & \multicolumn{1}{c}{\textbf{Description}}\\
\hline
1      & ENS   & 24  & 1205 & ENS domain name auction\\
2--3   & CMSW  & 49  & 503  & ConsenSys multisig wallet\\
4--5   & GMSW  & 49  & 704  & Gnosis multisig wallet\\
6      & BAT   & 23  & 191  & BAT token (advertising)\\
7      & CT    & 12  & 200  & ConsenSys token library\\
8      & ERCF  & 19  & 747  & ERC Fund (investment fund)\\
9      & FBT   & 34  & 385  & FirstBlood token (e-sports)\\
10--13 & HPN   & 173 & 3065 & Havven payment network\\
14     & MR    & 25  & 1053 & MicroRaiden payment service\\
15     & MT    & 38  & 437  & MOD token (supply-chain)\\
16     & PC    & 7   & 69   & Payment channel\\
17--18 & RNTS  & 49  & 749  & Request Network token sale\\
19     & DAO   & 23  & 783  & The DAO organization\\
20     & VT    & 18  & 242  & Valid token (personal data)\\
21     & USCC1 & 4   & 57   & USCC'17 entry\\
22     & USCC2 & 14  & 89   & USCC'17 (honorable mention)\\
23     & USCC3 & 21  & 535  & USCC'17 (3rd place)\\
24     & USCC4 & 7   & 164  & USCC'17 (1st place)\\
25     & USCC5 & 10  & 188  & USCC'17 (2nd place)\\
26     & PW    & 19  & 549  & Parity multisig wallet\\
27     & BNK    & 44  & 649  & Bankera token\\
\hline
\multicolumn{2}{c|}{\textbf{Total}} & 662 & \multicolumn{1}{r}{12564} &
\end{tabular}
}
\caption{Overview of benchmarks. The third and fourth columns provide
  the number of public functions and lines of source code in each
  benchmark, respectively.}
\label{tab:benchmarks}
\end{table}

\subsection{Experimental Setup}
\label{subsect:setup}

Our experiments compare the integration of \harvey and \bran
(incl. three variants) with \harvey alone to evaluate the
effectiveness of targeted fuzzing. The comparison focuses on the time
it takes for each configuration to cover a set of target locations.

\textbf{Targets.} We randomly selected up to four target locations for
each benchmark. In particular, we picked contract locations of varying
difficulty to reach, based on when they were first discovered during a
1h standard greybox-fuzzing run. So, we randomly picked at most one
newly discovered location, if one existed, from each of the following
time brackets in this order: 30--60m, 15--30m, 7.5--15m, 3.75--7.5m,
and 1.875--3.75m.

\textbf{Runs.} We performed 24 runs of each configuration on the 27
benchmarks of Tab.~\ref{tab:benchmarks}. For each run, we used a
different randomness seed, the same seed input, and a time limit of 1h
(i.e., 3'600s). In our results, we report medians and use
Wilcoxon-Mann-Whitney U tests to determine if differences in medians
between configurations are statistically significant.

\textbf{Machine.} We used an
Intel\textregistered~Xeon\textregistered~CPU~@~2.67GHz 24-core machine
with 50GB of memory running Debian 9.5.

\subsection{Results}
\label{subsect:results}




We now evaluate the effectiveness of our technique by investigating
five research questions.

\textbf{RQ1: Effectiveness of targeted fuzzing.}
Tab.~\ref{tab:resultsAvsB} compares our baseline configuration A,
which does not enable the static lookahead analysis, with
configuration B, which does. Note that configuration A uses the
cut-off-exponential power schedule of AFLFast~\cite{BoehmePham2016},
whereas B uses our specialized schedule. The first two columns of the
table indicate the benchmark and target IDs. Columns 3 and 4 show the
median time (in seconds) required to discover the first input that
reaches the target location (time-to-target) for both configurations,
and column 5 shows the speed-up factor. Column 6 shows the p-value,
which indicates the level of statistical significance; here, we use $p
< 0.05$ for ``significant'' differences. The last two columns show
Vargha-Delaney A12 effect sizes~\cite{VarghaDelaney2000}. Intuitively,
these measure the probability that configuration A is faster than B
and vice versa.

For 32 (out of 60) target locations, we observe significant
differences in time (i.e., $p < 0.05$), marked in bold in the
table. \emph{Configuration B significantly outperforms A for 31 (out
  of 32) of these target locations, with a median speed-up of up to
  14x} for one of the targets in benchmark 26.
In general, the results suggest that targeted fuzzing is very
effective, and unsurprisingly, its impact is most significant for
difficult targets (i.e., with high time-to-target for configuration
A). Specifically, \emph{for the 24 targets with $T_A \ge 900$ or $T_B
  \ge 900$, configuration B is significantly faster for 20, with
  insignificant differences between A and B for the remaining 4
  targets.}

Note that running the static analysis with an empty prefix (resembling
an offline analysis) on these benchmarks is not able to guide the
fuzzer at all. Since all our target locations are reachable by
construction, the analysis soundly reports them as
reachable. Therefore, the fuzzer still needs to explore the entire
contract to see if they indeed are.

\textbf{RQ2: Effectiveness of lookahead analysis.}
To measure the effect of the lookahead analysis, we created
configuration C, which is identical to configuration B except that the
analysis is maximally imprecise and inexpensive. Specifically,
$\Fcall{AreTargetsUnreachable}$ from Alg.~\ref{alg:LookaheadAnalysis}
simply returns false, and consequently, the computed $\mathit{LIDs}$
capture entire program paths, similarly to $\mathit{PIDs}$.

As shown in Tab.~\ref{tab:resultsCvsB}, there are significant
differences between configurations B and C for 21 target locations.
\emph{Configuration B is significantly faster than C for 17 out of 21
  targets}, and they are equally fast for 2 of the remaining 4
target locations.

Interestingly, configuration C is faster than A (for all 12 target
locations with significant differences). This suggests that our power
schedule regarding rare split points is effective independently of the
lookahead analysis.

\begin{table}[t!]
\centering
\scalebox{0.84}{
\begin{tabular}{r|r|r|r|r|r|r|r}
\multicolumn{1}{c|}{\textbf{BID}} & \multicolumn{1}{c|}{\textbf{Target ID}} & \multicolumn{1}{c|}{$\text{\textbf{T}}_{\text{\textbf{A}}}$} & \multicolumn{1}{c|}{$\text{\textbf{T}}_{\text{\textbf{B}}}$} & \multicolumn{1}{c|}{$\text{\textbf{T}}_{\text{\textbf{A}}}/\text{\textbf{T}}_{\text{\textbf{B}}}$} & \multicolumn{1}{c|}{$\text{\textbf{p}}$} & \multicolumn{1}{c|}{$\text{\textbf{A12}}_{\text{\textbf{A}}}$} & \multicolumn{1}{c}{$\text{\textbf{A12}}_{\text{\textbf{B}}}$}\\
\hline
1  & 79145a51:35ee & 324.15  & \textbf{90.25}   & 3.59  & 0.049 & 0.33 & 0.67\\
1  & 79145a51:bd4  & 32.69   & 69.53   & 0.47  & 0.130 & 0.63 & 0.37\\
2  & 060a46c9:d03  & 3385.55 & \textbf{706.71}  & 4.79  & 0.000 & 0.20 & 0.80\\
2  & 060a46c9:e29  & 161.66  & 106.57  & 1.52  & 0.197 & 0.39 & 0.61\\
2  & 060a46c9:16a5 & 701.39  & \textbf{339.86}  & 2.06  & 0.008 & 0.27 & 0.73\\
2  & 060a46c9:1f11 & 346.06  & \textbf{63.14}   & 5.48  & 0.000 & 0.11 & 0.89\\
3  & 708721b5:1485 & 396.11  & 394.54  & 1.00  & 0.477 & 0.44 & 0.56\\
3  & 708721b5:4ac  & 2292.00 & \textbf{775.93}  & 2.95  & 0.000 & 0.19 & 0.81\\
3  & 708721b5:1ca0 & 1248.59 & \textbf{817.76}  & 1.53  & 0.005 & 0.26 & 0.74\\
3  & 708721b5:1132 & 413.00  & \textbf{216.72}  & 1.91  & 0.003 & 0.24 & 0.76\\
4  & 9b8e6b2a:d08  & 3600.00 & \textbf{867.65}  & 4.15  & 0.000 & 0.15 & 0.85\\
4  & 9b8e6b2a:18f0 & 1657.33 & \textbf{432.50}  & 3.83  & 0.002 & 0.24 & 0.76\\
4  & 9b8e6b2a:1fee & 143.96  & 47.13   & 3.05  & 0.062 & 0.34 & 0.66\\
4  & 9b8e6b2a:553  & 3600.00 & \textbf{833.70}  & 4.32  & 0.001 & 0.22 & 0.78\\
5  & 5a3e5a7f:c09  & 3600.00 & \textbf{1282.42} & 2.81  & 0.000 & 0.08 & 0.92\\
5  & 5a3e5a7f:23f  & 900.53  & \textbf{466.99}  & 1.93  & 0.017 & 0.30 & 0.70\\
5  & 5a3e5a7f:1da8 & 1355.07 & \textbf{646.41}  & 2.10  & 0.000 & 0.16 & 0.84\\
5  & 5a3e5a7f:1d67 & 1497.96 & \textbf{524.08}  & 2.86  & 0.000 & 0.15 & 0.85\\
6  & 387bdf82:da7  & 61.66   & 22.70   & 2.72  & 0.089 & 0.36 & 0.64\\
8  & e2aedada:15a7 & 2592.56 & \textbf{1135.37} & 2.28  & 0.002 & 0.24 & 0.76\\
8  & e2aedada:17bb & 1783.03 & \textbf{612.39}  & 2.91  & 0.001 & 0.22 & 0.78\\
8  & e2aedada:d71  & 73.93   & 47.89   & 1.54  & 0.307 & 0.41 & 0.59\\
8  & e2aedada:13a8 & 258.14  & \textbf{74.87}   & 3.45  & 0.035 & 0.32 & 0.68\\
9  & dada6ee2:1693 & 334.82  & \textbf{49.38}   & 6.78  & 0.000 & 0.13 & 0.87\\
9  & dada6ee2:bee  & 225.12  & \textbf{72.14}   & 3.12  & 0.000 & 0.19 & 0.81\\
9  & dada6ee2:90e  & 84.62   & 50.39   & 1.68  & 0.338 & 0.42 & 0.58\\
10 & d98d1d6b:1f10 & 1124.84 & \textbf{281.45}  & 4.00  & 0.004 & 0.26 & 0.74\\
10 & d98d1d6b:401a & 164.12  & 153.95  & 1.07  & 0.861 & 0.48 & 0.52\\
10 & d98d1d6b:3cdd & 1669.91 & 1817.05 & 0.92  & 0.729 & 0.53 & 0.47\\
10 & d98d1d6b:3ce8 & 3600.00 & 3600.00 & 1.00  & 0.713 & 0.47 & 0.53\\
11 & 3ae06fbe:34db & 3600.00 & 3600.00 & 1.00  & 0.105 & 0.38 & 0.62\\
11 & 3ae06fbe:3de2 & 150.22  & 81.77   & 1.84  & 0.557 & 0.45 & 0.55\\
11 & 3ae06fbe:3ef3 & 284.34  & 395.15  & 0.72  & 0.703 & 0.47 & 0.53\\
11 & 3ae06fbe:10b2 & 238.35  & 142.03  & 1.68  & 0.228 & 0.40 & 0.60\\
12 & 0203d94d:713  & 76.82   & 60.27   & 1.27  & 0.910 & 0.49 & 0.51\\
14 & b8c706d1:125e & 3600.00 & 3600.00 & 1.00  & 0.085 & 0.39 & 0.61\\
14 & b8c706d1:3479 & 290.73  & 299.26  & 0.97  & 0.861 & 0.52 & 0.48\\
14 & b8c706d1:2023 & 34.65   & 43.72   & 0.79  & 0.992 & 0.50 & 0.50\\
15 & 06ef1a9c:27ce & 3365.87 & \textbf{467.90}  & 7.19  & 0.000 & 0.10 & 0.90\\
15 & 06ef1a9c:b41  & 100.00  & 73.83   & 1.35  & 0.877 & 0.49 & 0.51\\
15 & 06ef1a9c:a16  & 71.00   & 39.46   & 1.80  & 0.106 & 0.36 & 0.64\\
17 & 1c57401c:ef1  & 186.24  & 218.20  & 0.85  & 0.101 & 0.64 & 0.36\\
17 & 1c57401c:558  & 45.72   & 111.38  & 0.41  & 0.130 & 0.63 & 0.37\\
18 & ac0bf5ee:15e4 & 1827.66 & \textbf{321.36}  & 5.69  & 0.000 & 0.12 & 0.88\\
18 & ac0bf5ee:171b & 176.36  & \textbf{48.04}   & 3.67  & 0.000 & 0.16 & 0.84\\
18 & ac0bf5ee:15e0 & 133.84  & \textbf{27.80}   & 4.81  & 0.001 & 0.22 & 0.78\\
18 & ac0bf5ee:70c  & \textbf{24.87}   & 61.47   & 0.40  & 0.036 & 0.68 & 0.32\\
20 & 54142e12:1555 & 29.57   & 15.42   & 1.92  & 0.298 & 0.41 & 0.59\\
23 & d047b56e:5fb  & 42.01   & 20.70   & 2.03  & 0.279 & 0.41 & 0.59\\
24 & b9ebdb99:40c  & 980.79  & \textbf{139.78}  & 7.02  & 0.000 & 0.13 & 0.87\\
24 & b9ebdb99:3d1  & 282.28  & \textbf{57.21}   & 4.93  & 0.000 & 0.18 & 0.82\\
25 & f1e90f8f:9fd  & 316.48  & \textbf{24.61}   & 12.86 & 0.000 & 0.09 & 0.91\\
26 & a788e7af:1f07 & 1778.07 & \textbf{130.34}  & 13.64 & 0.000 & 0.07 & 0.93\\
26 & a788e7af:1e29 & 2005.67 & \textbf{336.04}  & 5.97  & 0.000 & 0.12 & 0.88\\
26 & a788e7af:544  & 395.22  & 47.84   & 8.26  & 0.140 & 0.38 & 0.62\\
26 & a788e7af:32b  & 44.67   & 45.92   & 0.97  & 0.813 & 0.48 & 0.52\\
27 & 9473c978:1541 & 2445.87 & \textbf{324.46}  & 7.54  & 0.020 & 0.31 & 0.69\\
27 & 9473c978:e33  & 1493.03 & \textbf{637.16}  & 2.34  & 0.023 & 0.31 & 0.69\\
27 & 9473c978:150e & 178.11  & 97.60   & 1.82  & 0.120 & 0.37 & 0.63\\
27 & 9473c978:8e8  & 102.29  & 150.72  & 0.68  & 0.236 & 0.60 & 0.40\\
\hline
\end{tabular}
}
\caption{Comparing time-to-target between configuration A (w/o lookahead analysis) and B (w/ lookahead analysis).}
\label{tab:resultsAvsB}
\vspace{-1em}
\end{table}

\textbf{RQ3: Effectiveness of power schedule.}
To measure the effect of targeting rare $\mathit{LIDs}$ and rare split
points in our power schedule, we created configuration D. It is
identical to configuration B except that it uses a variant of
AFLFast's cut-off-exponential power
schedule~\cite{BoehmePham2016}. The original power schedule assigns
energy to an input $I$ based on how often its $\mathit{PID}$ has been
exercised. In contrast, our variant is based on how often its
$\mathit{LID}$ has been exercised and corresponds to using the results
of the lookahead analysis with a standard power schedule.

However, as shown in Tab.~\ref{tab:resultsDvsB}, \emph{configuration B
  is faster than configuration D for 28 of 30 targets (with
  significant differences).}  This indicates that our power schedule
significantly reduces the time-to-target, thus effectively guiding the
fuzzer.

Nonetheless, configuration D is faster than A for all 6 targets with
significant differences. This shows the effectiveness of the
lookahead analysis independently of the power schedule.

\begin{table}[t!]
\centering
\scalebox{0.84}{
\begin{tabular}{r|r|r|r|r|r|r|r}
\multicolumn{1}{c|}{\textbf{BID}} & \multicolumn{1}{c|}{\textbf{Target ID}} & \multicolumn{1}{c|}{$\text{\textbf{T}}_{\text{\textbf{C}}}$} & \multicolumn{1}{c|}{$\text{\textbf{T}}_{\text{\textbf{B}}}$} & \multicolumn{1}{c|}{$\text{\textbf{T}}_{\text{\textbf{C}}}/\text{\textbf{T}}_{\text{\textbf{B}}}$} & \multicolumn{1}{c|}{$\text{\textbf{p}}$} & \multicolumn{1}{c|}{$\text{\textbf{A12}}_{\text{\textbf{A}}}$} & \multicolumn{1}{c}{$\text{\textbf{A12}}_{\text{\textbf{B}}}$}\\
\hline
1 &  79145a51:35ee & 223.45  & 90.25   & 2.48 & 0.718 & 0.47 & 0.53\\
1 &  79145a51:bd4  & 69.07   & 69.53   & 0.99 & 0.658 & 0.46 & 0.54\\
2 &  060a46c9:d03  & 2164.66 & \textbf{706.71}  & 3.06 & 0.005 & 0.27 & 0.73\\
2 &  060a46c9:e29  & 156.18  & 106.57  & 1.47 & 0.338 & 0.42 & 0.58\\
2 &  060a46c9:16a5 & 854.32  & \textbf{339.86}  & 2.51 & 0.042 & 0.33 & 0.67\\
2 &  060a46c9:1f11 & 56.01   & 63.14   & 0.89 & 0.926 & 0.49 & 0.51\\
3 &  708721b5:1485 & 527.02  & 394.54  & 1.34 & 0.797 & 0.48 & 0.52\\
3 &  708721b5:4ac  & 2000.32 & \textbf{775.93}  & 2.58 & 0.007 & 0.27 & 0.73\\
3 &  708721b5:1ca0 & 775.97  & 817.76  & 0.95 & 0.327 & 0.42 & 0.58\\
3 &  708721b5:1132 & 298.71  & 216.72  & 1.38 & 0.317 & 0.41 & 0.59\\
4 &  9b8e6b2a:d08  & 3600.00 & \textbf{867.65}  & 4.15 & 0.000 & 0.07 & 0.93\\
4 &  9b8e6b2a:18f0 & 1288.76 & \textbf{432.50}  & 2.98 & 0.008 & 0.28 & 0.72\\
4 &  9b8e6b2a:1fee & 88.80   & 47.13   & 1.88 & 0.557 & 0.45 & 0.55\\
4 &  9b8e6b2a:553  & 3508.27 & \textbf{833.70}  & 4.21 & 0.000 & 0.10 & 0.90\\
5 &  5a3e5a7f:c09  & 3600.00 & \textbf{1282.42} & 2.81 & 0.000 & 0.09 & 0.91\\
5 &  5a3e5a7f:23f  & 2102.80 & \textbf{466.99}  & 4.50 & 0.000 & 0.19 & 0.81\\
5 &  5a3e5a7f:1da8 & 1961.40 & \textbf{646.41}  & 3.03 & 0.001 & 0.21 & 0.79\\
5 &  5a3e5a7f:1d67 & 1977.32 & \textbf{524.08}  & 3.77 & 0.001 & 0.22 & 0.78\\
6 &  387bdf82:da7  & 20.35   & 22.70   & 0.90 & 0.317 & 0.59 & 0.41\\
8 &  e2aedada:15a7 & 2381.94 & \textbf{1135.37} & 2.10 & 0.004 & 0.26 & 0.74\\
8 &  e2aedada:17bb & 915.16  & 612.39  & 1.49 & 0.071 & 0.35 & 0.65\\
8 &  e2aedada:d71  & 30.51   & 47.89   & 0.64 & 0.571 & 0.55 & 0.45\\
8 &  e2aedada:13a8 & 91.11   & 74.87   & 1.22 & 0.845 & 0.48 & 0.52\\
9 &  dada6ee2:1693 & 253.55  & \textbf{49.38}   & 5.13 & 0.000 & 0.18 & 0.82\\
9 &  dada6ee2:bee  & 111.31  & \textbf{72.14}   & 1.54 & 0.038 & 0.32 & 0.68\\
9 &  dada6ee2:90e  & 49.37   & 50.39   & 0.98 & 0.628 & 0.54 & 0.46\\
10 & d98d1d6b:1f10 & 139.34  & 281.45  & 0.50 & 0.093 & 0.64 & 0.36\\
10 & d98d1d6b:401a & 145.53  & 153.95  & 0.95 & 0.829 & 0.52 & 0.48\\
10 & d98d1d6b:3ce8 & 3600.00 & 3600.00 & 1.00 & 0.226 & 0.41 & 0.59\\
10 & d98d1d6b:3cdd & 3600.00 & 1817.05 & 1.98 & 0.146 & 0.38 & 0.62\\
11 & 3ae06fbe:34db & \textbf{3600.00} & \textbf{3600.00} & 1.00 & 0.027 & 0.35 & 0.65\\
11 & 3ae06fbe:3de2 & 169.72  & 81.77   & 2.08 & 0.158 & 0.38 & 0.62\\
11 & 3ae06fbe:3ef3 & 182.77  & 395.15  & 0.46 & 0.135 & 0.63 & 0.37\\
11 & 3ae06fbe:10b2 & 214.12  & 142.03  & 1.51 & 0.942 & 0.51 & 0.49\\
12 & 0203d94d:713  & 44.13   & 60.27   & 0.73 & 0.516 & 0.56 & 0.44\\
14 & b8c706d1:125e & \textbf{3600.00} & \textbf{3600.00} & 1.00 & 0.010 & 0.35 & 0.65\\
14 & b8c706d1:3479 & 108.30  & 299.26  & 0.36 & 0.110 & 0.64 & 0.36\\
14 & b8c706d1:2023 & 40.71   & 43.72   & 0.93 & 0.845 & 0.52 & 0.48\\
15 & 06ef1a9c:27ce & 2458.74 & \textbf{467.90}  & 5.25 & 0.001 & 0.23 & 0.77\\
15 & 06ef1a9c:b41  & 59.20   & 73.83   & 0.80 & 0.228 & 0.60 & 0.40\\
15 & 06ef1a9c:a16  & 57.15   & 39.46   & 1.45 & 0.529 & 0.45 & 0.55\\
17 & 1c57401c:ef1  & \textbf{104.23}  & 218.20  & 0.48 & 0.009 & 0.72 & 0.28\\
17 & 1c57401c:558  & \textbf{63.84}   & 111.38  & 0.57 & 0.009 & 0.72 & 0.28\\
18 & ac0bf5ee:15e4 & 719.04  & \textbf{321.36}  & 2.24 & 0.007 & 0.27 & 0.73\\
18 & ac0bf5ee:171b & 106.78  & \textbf{48.04}   & 2.22 & 0.002 & 0.23 & 0.77\\
18 & ac0bf5ee:15e0 & 21.29   & 27.80   & 0.77 & 0.370 & 0.58 & 0.42\\
18 & ac0bf5ee:70c  & 26.28   & 61.47   & 0.43 & 0.051 & 0.66 & 0.34\\
20 & 54142e12:1555 & 17.67   & 15.42   & 1.15 & 0.585 & 0.55 & 0.45\\
23 & d047b56e:5fb  & 17.53   & 20.70   & 0.85 & 0.571 & 0.55 & 0.45\\
24 & b9ebdb99:40c  & 178.49  & 139.78  & 1.28 & 0.138 & 0.37 & 0.63\\
24 & b9ebdb99:3d1  & 115.03  & 57.21   & 2.01 & 0.089 & 0.36 & 0.64\\
25 & f1e90f8f:9fd  & 114.00  & \textbf{24.61}   & 4.63 & 0.000 & 0.16 & 0.84\\
26 & a788e7af:1f07 & 323.97  & 130.34  & 2.49 & 0.089 & 0.36 & 0.64\\
26 & a788e7af:1e29 & 404.19  & 336.04  & 1.20 & 0.797 & 0.48 & 0.52\\
26 & a788e7af:544  & 142.41  & 47.84   & 2.98 & 0.464 & 0.44 & 0.56\\
26 & a788e7af:32b  & 40.09   & 45.92   & 0.87 & 0.992 & 0.50 & 0.50\\
27 & 9473c978:1541 & 2320.70 & 324.46  & 7.15 & 0.210 & 0.39 & 0.61\\
27 & 9473c978:e33  & 1824.92 & 637.16  & 2.86 & 0.052 & 0.34 & 0.66\\
27 & 9473c978:150e & 49.45   & 97.60   & 0.51 & 0.205 & 0.61 & 0.39\\
27 & 9473c978:8e8  & 95.71   & 150.72  & 0.63 & 0.244 & 0.60 & 0.40\\
\hline
\end{tabular}
}
\caption{Comparing time-to-target for configurations B and C.}
\label{tab:resultsCvsB}
\vspace{-1em}
\end{table}

\textbf{RQ4: Scalability of lookahead analysis.}
One key design decision for using an \emph{online} static analysis as part of a dynamic
analysis (i.e., greybox fuzzing) was to make the static analysis as lightweight and
scalable as sensible. That is why our lookahead analysis is modular and uses an inexpensive
constant-propagation domain. 

Our results confirm that \emph{the running time of the lookahead
  analysis is a tiny fraction of the total running time of the fuzzer
  (0.09--105.93s of a total of 3600s per benchmark, median 2.73s).}
This confirms that even a very lightweight static analysis can boost
the effectiveness of fuzzing.

\textbf{RQ5: Effect on instruction coverage.}
In our evaluation, \emph{there were no noticeable instruction-coverage differences
  between any of our configurations.}

This indicates that our approach to targeted greybox fuzzing mainly
affects the order in which different program locations are
reached. Even though we prioritize certain inputs by assigning more
energy to them, the fuzzer still mutates them randomly and eventually
covers the same instructions as standard fuzzing. To avoid this, we
would need to restrict some mutations (e.g., ones that never discover
new $\mathit{LIDs}$), much like FairFuzz~\cite{LemieuxSen2018}
restricts mutations that do not reach rare branches.

\begin{table}[t!]
\centering
\scalebox{0.84}{
\begin{tabular}{r|r|r|r|r|r|r|r}
\multicolumn{1}{c|}{\textbf{BID}} & \multicolumn{1}{c|}{\textbf{Target ID}} & \multicolumn{1}{c|}{$\text{\textbf{T}}_{\text{\textbf{D}}}$} & \multicolumn{1}{c|}{$\text{\textbf{T}}_{\text{\textbf{B}}}$} & \multicolumn{1}{c|}{$\text{\textbf{T}}_{\text{\textbf{D}}}/\text{\textbf{T}}_{\text{\textbf{B}}}$} & \multicolumn{1}{c|}{$\text{\textbf{p}}$} & \multicolumn{1}{c|}{$\text{\textbf{A12}}_{\text{\textbf{A}}}$} & \multicolumn{1}{c}{$\text{\textbf{A12}}_{\text{\textbf{B}}}$}\\
\hline
1  & 79145a51:35ee & 252.95  & \textbf{90.25}   & 2.80 & 0.030 & 0.32 & 0.68\\
1  & 79145a51:bd4  & 64.12   & 69.53   & 0.92 & 0.688 & 0.53 & 0.47\\
2  & 060a46c9:d03  & 1734.13 & \textbf{706.71}  & 2.45 & 0.013 & 0.29 & 0.71\\
2  & 060a46c9:e29  & 246.00  & \textbf{106.57}  & 2.31 & 0.042 & 0.33 & 0.67\\
2  & 060a46c9:16a5 & 579.02  & 339.86  & 1.70 & 0.120 & 0.37 & 0.63\\
2  & 060a46c9:1f11 & 219.87  & \textbf{63.14}   & 3.48 & 0.000 & 0.19 & 0.81\\
3  & 708721b5:1485 & 337.42  & 394.54  & 0.86 & 0.781 & 0.52 & 0.48\\
3  & 708721b5:4ac  & 1553.51 & \textbf{775.93}  & 2.00 & 0.013 & 0.29 & 0.71\\
3  & 708721b5:1ca0 & 1001.05 & 817.76  & 1.22 & 0.183 & 0.39 & 0.61\\
3  & 708721b5:1132 & 353.12  & \textbf{216.72}  & 1.63 & 0.049 & 0.33 & 0.67\\
4  & 9b8e6b2a:d08  & 1353.86 & \textbf{867.65}  & 1.56 & 0.030 & 0.32 & 0.68\\
4  & 9b8e6b2a:18f0 & 1008.23 & \textbf{432.50}  & 2.33 & 0.033 & 0.32 & 0.68\\
4  & 9b8e6b2a:1fee & 172.58  & \textbf{47.13}   & 3.66 & 0.002 & 0.24 & 0.76\\
4  & 9b8e6b2a:553  & 2464.13 & \textbf{833.70}  & 2.96 & 0.000 & 0.16 & 0.84\\
5  & 5a3e5a7f:c09  & 3381.14 & \textbf{1282.42} & 2.64 & 0.001 & 0.21 & 0.79\\
5  & 5a3e5a7f:23f  & 515.76  & 466.99  & 1.10 & 0.220 & 0.40 & 0.60\\
5  & 5a3e5a7f:1da8 & 1197.92 & \textbf{646.41}  & 1.85 & 0.002 & 0.24 & 0.76\\
5  & 5a3e5a7f:1d67 & 855.79  & \textbf{524.08}  & 1.63 & 0.003 & 0.25 & 0.75\\
6  & 387bdf82:da7  & 110.41  & \textbf{22.70}   & 4.86 & 0.000 & 0.18 & 0.82\\
8  & e2aedada:15a7 & 2194.73 & \textbf{1135.37} & 1.93 & 0.002 & 0.24 & 0.76\\
8  & e2aedada:17bb & 1021.35 & 612.39  & 1.67 & 0.101 & 0.36 & 0.64\\
8  & e2aedada:d71  & 82.30   & 47.89   & 1.72 & 0.097 & 0.36 & 0.64\\
8  & e2aedada:13a8 & 188.01  & 74.87   & 2.51 & 0.051 & 0.34 & 0.66\\
9  & dada6ee2:1693 & 279.31  & \textbf{49.38}   & 5.66 & 0.001 & 0.23 & 0.77\\
9  & dada6ee2:bee  & 195.79  & \textbf{72.14}   & 2.71 & 0.006 & 0.27 & 0.73\\
9  & dada6ee2:90e  & 45.93   & 50.39   & 0.91 & 0.992 & 0.50 & 0.50\\
10 & d98d1d6b:1f10 & 606.63  & 281.45  & 2.16 & 0.085 & 0.35 & 0.65\\
10 & d98d1d6b:3ce8 & 3600.00 & 3600.00 & 1.00 & 0.840 & 0.52 & 0.48\\
10 & d98d1d6b:401a & 254.15  & 153.95  & 1.65 & 0.228 & 0.40 & 0.60\\
10 & d98d1d6b:3cdd & 1956.69 & 1817.05 & 1.08 & 0.857 & 0.48 & 0.52\\
11 & 3ae06fbe:34db & 3591.91 & 3600.00 & 1.00 & 0.885 & 0.51 & 0.49\\
11 & 3ae06fbe:3de2 & 181.38  & 81.77   & 2.22 & 0.130 & 0.37 & 0.63\\
11 & 3ae06fbe:3ef3 & 383.75  & 395.15  & 0.97 & 0.158 & 0.38 & 0.62\\
11 & 3ae06fbe:10b2 & 163.65  & 142.03  & 1.15 & 0.781 & 0.48 & 0.52\\
12 & 0203d94d:713  & 38.85   & 60.27   & 0.64 & 0.220 & 0.60 & 0.40\\
14 & b8c706d1:125e & 3600.00 & 3600.00 & 1.00 & 0.449 & 0.45 & 0.55\\
14 & b8c706d1:3479 & 501.51  & 299.26  & 1.68 & 0.338 & 0.42 & 0.58\\
14 & b8c706d1:2023 & 62.22   & 43.72   & 1.42 & 0.164 & 0.38 & 0.62\\
15 & 06ef1a9c:27ce & 2514.11 & \textbf{467.90}  & 5.37 & 0.000 & 0.10 & 0.90\\
15 & 06ef1a9c:b41  & 119.89  & 73.83   & 1.62 & 0.252 & 0.40 & 0.60\\
15 & 06ef1a9c:a16  & 102.73  & \textbf{39.46}   & 2.60 & 0.020 & 0.30 & 0.70\\
17 & 1c57401c:ef1  & \textbf{89.83}   & 218.20  & 0.41 & 0.025 & 0.69 & 0.31\\
17 & 1c57401c:558  & 66.72   & 111.38  & 0.60 & 0.184 & 0.61 & 0.39\\
18 & ac0bf5ee:15e4 & 947.01  & \textbf{321.36}  & 2.95 & 0.020 & 0.30 & 0.70\\
18 & ac0bf5ee:171b & 177.27  & \textbf{48.04}   & 3.69 & 0.004 & 0.25 & 0.75\\
18 & ac0bf5ee:15e0 & 72.29   & 27.80   & 2.60 & 0.071 & 0.35 & 0.65\\
18 & ac0bf5ee:70c  & \textbf{29.28}   & 61.47   & 0.48 & 0.021 & 0.69 & 0.31\\
20 & 54142e12:1555 & 24.46   & 15.42   & 1.59 & 0.516 & 0.44 & 0.56\\
23 & d047b56e:5fb  & 36.38   & 20.70   & 1.76 & 0.348 & 0.42 & 0.58\\
24 & b9ebdb99:40c  & 785.68  & \textbf{139.78}  & 5.62 & 0.000 & 0.15 & 0.85\\
24 & b9ebdb99:3d1  & 221.02  & \textbf{57.21}   & 3.86 & 0.000 & 0.15 & 0.85\\
25 & f1e90f8f:9fd  & 232.58  & \textbf{24.61}   & 9.45 & 0.000 & 0.01 & 0.99\\
26 & a788e7af:1f07 & 533.02  & \textbf{130.34}  & 4.09 & 0.016 & 0.30 & 0.70\\
26 & a788e7af:1e29 & 513.20  & 336.04  & 1.53 & 0.599 & 0.45 & 0.55\\
26 & a788e7af:544  & 335.21  & \textbf{47.84}   & 7.01 & 0.028 & 0.31 & 0.69\\
26 & a788e7af:32b  & 72.62   & 45.92   & 1.58 & 0.543 & 0.45 & 0.55\\
27 & 9473c978:1541 & 1938.89 & \textbf{324.46}  & 5.98 & 0.027 & 0.31 & 0.69\\
27 & 9473c978:e33  & 1517.21 & \textbf{637.16}  & 2.38 & 0.024 & 0.31 & 0.69\\
27 & 9473c978:150e & 160.07  & 97.60   & 1.64 & 0.093 & 0.36 & 0.64\\
27 & 9473c978:8e8  & 112.27  & 150.72  & 0.74 & 0.543 & 0.55 & 0.45\\
\hline
\end{tabular}
}
\caption{Comparing time-to-target for configurations B and D.}
\label{tab:resultsDvsB}
\vspace{-1em}
\end{table}

\subsection{Threats to Validity}
\label{subsect:threats}

We have identified the following threats to validity.

\textbf{External validity.} A potential threat to the validity of our
experiments has to do with external
validity~\cite{SiegmundSiegmund2015}. In particular, our results may
not generalize to other contracts or programs. To alleviate this
threat, we selected benchmarks from several, diverse application
domains. Moreover, in the appendix, we provide the
versions of all contracts used in our experiments so that others can
also test them. The results may also not generalize to other target
locations, but we alleviate this threat by selecting them at random
and with varying difficulty to reach.

\textbf{Internal validity.} Internal
validity~\cite{SiegmundSiegmund2015} is compromised when systematic
errors are introduced in the experimental setup. A common pitfall in
evaluating randomized approaches, such as fuzzing, is the potentially
biased selection of seeds. During our experiments, when comparing the
different configurations of our technique, we consistently used the
same seed inputs for \harvey.

\textbf{Construct validity.} Construct validity ensures that any
improvements, for instance in effectiveness or performance, achieved
by a particular technique are due to that technique alone, and not due
to other factors, such as better engineering. In our experiments, we
compare different configurations of the same greybox fuzzer, and
consequently, any effect on the results is exclusively caused by their
differences.

\section{Related Work}
\label{sect:relatedWork}

Our technique for targeted greybox fuzzing leverages an online static
analysis to semantically analyze each new path that is added to the
fuzzer's test suite. The feedback collected by the static analysis is
used to guide the fuzzer toward a set of target locations using a
novel power schedule that takes inspiration from two existing
ones~\cite{BoehmePham2016, LemieuxSen2018}.

In contrast, the most closely related work~\cite{BoehmePham2017}
performs an offline instrumentation of the program under test encoding
a static distance metric between the instrumented and the target
locations in the control-flow graph. When running a given input, the
instrumentation is used to obtain a dynamic (aggregated)
distance. This distance subsequently guides the fuzzer toward the
target locations.

Since a control-flow graph cannot always be easily recovered from EVM
bytecode (e.g., due to indirect jumps), our lookahead analysis
directly analyzes the bytecode using abstract
interpretation~\cite{CousotCousot1977,CousotCousot1979}. Our
implementation uses the constant-propagation domain~\cite{Kildall1973}
to track the current state of the EVM (for instance, to resolve jump
targets that are pushed to the execution stack). Unlike traditional
static analyses, it aims to improve precision by performing a
partially path-sensitive analysis---that is, path-sensitive for a
prefix of a feasible path recorded at runtime by the fuzzer, and
path-insensitive for all suffix paths.

\textbf{Guiding greybox fuzzers.}
Besides directed greybox fuzzing~\cite{BoehmePham2017}, there is a
number of greybox fuzzers that target specific program
locations~\cite{ChenXue2018}, rare branches~\cite{LemieuxSen2018},
uncovered branches~\cite{LiChen2017,WangLiang2018}, or suspected
vulnerabilities~\cite{GaneshLeek2009,HallerSlowinska2013,LiJi2019,ChowdhuryMedicherla2019}. While
several of these fuzzers use an offline static analysis to guide the
exploration, none of them leverages an online analysis.

\textbf{Guiding other program analyzers.}
There is a large body of work on guiding analyzers toward specific
target locations~\cite{MaKhoo2011,MarinescuCadar2013} or potential
failures~\cite{CsallnerSmaragdakis2005,DwyerPurandare2007,NoriRajamani2009,GodefroidNori2010,GeTaneja2011,CzechJakobs2015,MaArtho2015,ChristakisMueller2016,FerlesWuestholz2017,DevecseryChen2018}
by combining static and dynamic analysis. These combinations typically
perform an offline static analysis first and use it to improve the
effectiveness of a subsequent dynamic analysis; for instance, by
pruning parts of the program.
For example, Check 'n' Crash~\cite{CsallnerSmaragdakis2005} integrates
the ESC/Java static checker~\cite{FlanaganLeino2002} with the JCrasher
test-generation tool~\cite{CsallnerSmaragdakis2004}. Similarly,
DyTa~\cite{GeTaneja2011} combines the .NET static analyzer
Clousot~\cite{FahndrichLogozzo2010} with the dynamic symbolic
execution engine
Pex~\cite{TillmanndeHalleux2008}. YOGI~\cite{NoriRajamani2009,GodefroidNori2010}
constantly refines its over- and under-approximations in the style of
counterexample-driven refinement~\cite{ClarkeGrumberg2000}.
In contrast, our lookahead analysis is online and constitutes a core
component of our targeted greybox fuzzer.

Hybrid concolic testing~\cite{MajumdarSen2007} combines random testing
with concolic
testing~\cite{GodefroidKlarlund2005,CadarEngler2005,SenAgha2006}. Even
though the technique significantly differs from ours, it shares an
interesting similarity: it uses online concolic testing during a
concrete program execution to discover uncovered code on-the-fly. When
successful, the inputs for covering the code are used to resume the
concrete program execution.

\textbf{Symbolic execution.}
In the context of symbolic execution~\cite{King1976}, there have
emerged numerous search strategies for guiding the exploration; for
instance, to target deeper paths (in depth-first search), uncovered
statements~\cite{ParkHossain2012}, or ``less-traveled
paths''~\cite{LiSu2013}.
Our technique resembles a search strategy in that it prioritizes
exploration of certain inputs over others.

Compositional symbolic
execution~\cite{Godefroid2007,AnandGodefroid2008} has been shown to be
effective in merging different program paths by means of summaries in
order to alleviate path explosion. Dynamic state
merging~\cite{KuznetsovKinder2012} and
veritesting~\cite{AvgerinosRebert2014} can also be seen as forms of
summarization. Similarly, our technique merges different paths that
share the same lookahead identifier for the purpose of assigning
energy. The more precise the lookahead analysis, the shorter the
no-target-ahead prefixes, and thus, the more effective the merging.

\textbf{Program analysis for smart contracts.}
There is a growing number of program analyzers for smart contracts, ranging from random
test generation frameworks to static analyzers and
verifiers~\cite{LuuChu2016,BhargavanDelignat-Lavaud2016,AtzeiBartoletti2017,ChenLi2017,SergeyHobor2017,JiangLiu2018,ChatterjeeGoharshady2018,AmaniBegel2018,BrentJurisevic2018,GrechKong2018,GrossmanAbraham2018,KalraGoel2018,NikolicKolluri2018,TsankovDan2018,Echidna,Manticore,Mythril}.
In contrast, we present a targeted greybox fuzzer for smart contracts,
the first analyzer for contracts that incorporates static and dynamic
analysis.

\section{Conclusion}
\label{sect:conclusion}

We have presented a novel technique for targeted fuzzing using static
lookahead analysis. The key idea is to enable a symbiotic
collaboration between the greybox fuzzer and an online static
analysis. On one hand, dynamic information (i.e., feasible program
paths) are used to boost the precision of the static analysis. On the
other hand, static information about reachable target locations---more
specifically, lookahead identifiers and split points---is used to
guide the greybox fuzzer toward target locations. Our experiments on
27 real-world benchmarks show that targeted fuzzing significantly
outperforms standard greybox fuzzing for reaching 83\% of the
challenging target locations (up to 14x of median speed-up).

In future work, we plan to investigate other combinations of dynamic
and online static analysis; for instance, to guide dynamic symbolic
execution.

\newpage

\bibliographystyle{IEEEtran}
\bibliography{tandem}

\newpage

\onecolumn

\appendix

\section{Smart Contract Repositories}
\label{sect:repos}

All tested smart contracts are open source. Tab.~\ref{tab:repos}
provides the changeset IDs and links to their repositories.

\begin{table*}[b!]
\centering
\scalebox{0.9}{
\begin{tabular}{r|l|l|l}
\multicolumn{1}{c|}{\textbf{BIDs}} & \multicolumn{1}{c|}{\textbf{Name}} & \multicolumn{1}{c|}{\textbf{Changeset ID}} & \multicolumn{1}{c}{\textbf{Repository}}\\
\hline
1      & ENS   & 5108f51d656f201dc0054e55f5fd000d00ef9ef3 & \url{https://github.com/ethereum/ens}\\
2--3   & CMSW  & 2582787a14dd861b51df6f815fab122ff51fb574 & \url{https://github.com/ConsenSys/MultiSigWallet}\\
4--5   & GMSW  & 8ac8ba7effe6c3845719e480defb5f2ecafd2fd4 & \url{https://github.com/gnosis/MultiSigWallet}\\
6      & BAT   & 15bebdc0642dac614d56709477c7c31d5c993ae1 & \url{https://github.com/brave-intl/basic-attention-token-crowdsale}\\
7      & CT    & 1f62e1ba3bf32dc22fe2de94a9ee486d667edef2 & \url{https://github.com/ConsenSys/Tokens}\\
8      & ERCF  & c7d025220a1388326b926d8983e47184e249d979 & \url{https://github.com/ScJa/ercfund}\\
9      & FBT   & ae71053e0656b0ceba7e229e1d67c09f271191dc & \url{https://github.com/Firstbloodio/token}\\
10--13 & HPN   & 540006e0e2e5ef729482ad8bebcf7eafcd5198c2 & \url{https://github.com/Havven/havven}\\
14     & MR    & 527eb90c614ff4178b269d48ea063eb49ee0f254 & \url{https://github.com/raiden-network/microraiden}\\
15     & MT    & 7009cc95affa5a2a41a013b85903b14602c25b4f & \url{https://github.com/modum-io/tokenapp-smartcontract}\\
16     & PC    & 515c1b935ac43afc6bf54fcaff68cf8521595b0b & \url{https://github.com/mattdf/payment-channel}\\
17--18 & RNTS  & 6c39082eff65b2d3035a89a3f3dd94bde6cca60f & \url{https://github.com/RequestNetwork/RequestTokenSale}\\
19     & DAO   & f347c0e177edcfd99d64fe589d236754fa375658 & \url{https://github.com/slockit/DAO}\\
20     & VT    & 30ede971bb682f245e5be11f544e305ef033a765 & \url{https://github.com/valid-global/token}\\
21     & USCC1 & 3b26643a85d182a9b8f0b6fe8c1153f3bd510a96 & \url{https://github.com/Arachnid/uscc}\\
22     & USCC2 & 3b26643a85d182a9b8f0b6fe8c1153f3bd510a96 & \url{https://github.com/Arachnid/uscc}\\
23     & USCC3 & 3b26643a85d182a9b8f0b6fe8c1153f3bd510a96 & \url{https://github.com/Arachnid/uscc}\\
24     & USCC4 & 3b26643a85d182a9b8f0b6fe8c1153f3bd510a96 & \url{https://github.com/Arachnid/uscc}\\
25     & USCC5 & 3b26643a85d182a9b8f0b6fe8c1153f3bd510a96 & \url{https://github.com/Arachnid/uscc}\\
26     & PW    & 657da22245dcfe0fe1cccc58ee8cd86924d65cdd & \url{https://github.com/paritytech/contracts}\\
27     & BNK   & 97f1c3195bc6f4d8b3393016ecf106b42a2b1d97 & \url{https://github.com/Bankera-token/BNK-ETH-Contract}
\end{tabular}
}
\vspace{1em}
\caption{Smart contract repositories.}
\label{tab:repos}
\end{table*}

\end{document}